\documentclass[aps,pre,twocolumn,groupedaddress,showkeys]{revtex4}
\usepackage{float}
\usepackage{graphicx}
\usepackage{amstext}
\usepackage{amssymb}
\usepackage{amsmath}
\usepackage{dcolumn}
\usepackage{multirow}

\makeatletter
\renewenvironment{widetext@grid}{%
  \par\ignorespaces
  \setbox\widetext@top\vbox{%
   \vskip15\p@
   \hb@xt@\hsize{%
    \leaders\hrule\hfil
    \vrule\@height6\p@
   }%
   \vskip6\p@
  }%
  \setbox\widetext@bot\hb@xt@\hsize{%
    \vrule\@depth6\p@
    \leaders\hrule\hfil
  }%
  \onecolumngrid
  \let\set@footnotewidth\set@footnotewidth@ii
}{%
  \par
  \setbox\widetext@bot\vbox{%
   \hb@xt@\hsize{\hfil\box\widetext@bot}%
   \vskip14\p@
  }%
  \twocolumngrid\global\@ignoretrue
  \@endpetrue
}%
\renewcommand*\env@matrix[1][\arraystretch]{%
  \edef\arraystretch{#1}%
  \hskip -\arraycolsep
  \let\@ifnextchar\new@ifnextchar
  \array{*\c@MaxMatrixCols c}}%
\makeatother

\begin{document}

\title{Regime-Specific Multi-Objective Variational Principle\\of Compromise in Competition at Mesoscales}

\author{Jinghai Li}
\email[E-mail:]{jhli@ipe.ac.cn}
\author{Wenlai Huang}
\author{Jianhua Chen}
\author{Wei Ge}
\author{Chaofeng Hou}
\affiliation{State Key Laboratory of Multiphase Complex Systems, Institute of Process Engineering, Chinese Academy of Sciences, Beijing 100190, People's Republic of China}

\date{\today}

\begin{abstract}
The energy-minimization multiscale (EMMS) principle of compromise in competition is believed to be generally applicable for all mesoscale problems at different levels in the real world, spanning from elementary particles to the universe. This stimulated a fundamental proposition of the concept of mesoscience. This article discusses a potential universality of the regime-specific multi-objective variational feature of this underlying principle through case studies in chemical engineering. It is also elucidated why the currently available variational principles are not applicable to the mesoscale problems. The paper concludes with prospects on future study.
\end{abstract}


\maketitle

\section{Introduction}
\label{mysec1}
In the past decades, science has made remarkable progress in extending its reachable scales with respect to both space and time, and in shifting its paradigm and landscape of methodology and tools. However, while we know more and more details at smaller and smaller scales, such as understanding of genomics and elementary particles, we cannot yet fully reveal the secret of life and the relationship between material structures and their evident macroscopic properties. On the other hand, coarse-graining approaches are widely used at different levels of sciences to treat problems associated with complex structures, while science, devoted to bridging the microscale and macroscale - termed complexity science - is still in its infancy. It seems clear that some links have been missing in the context of contemporary science, leading to gaps of knowledge, which block the way to finding the solutions for many challenging issues.

Meanwhile, we recognized that the currently available theories are not applicable to the mesoscale dynamic structures which are likely governed by at least two dominant mechanisms \cite{NSR2017,AR2017}, leading to the multi-objective variational feature.

We believe that these gaps originated from a common cause; namely the missing of governing principles between the ``\emph{unit scales}" and ``\emph{system scales}" at different levels of the real world \cite{book1994}. We predicted that one of the approaches to remove these gaps in our knowledge is to resolve the corresponding mesoscale problems \cite{arXiv2009,P2010-634}, through the energy-minimization multiscale (EMMS) encompassing the underlying principle of ``Compromise in competition" \cite{book2013,book2014}. This principle is believed to be regime-specific, multi-objective variational, and possibly universal, hopefully yielding fundamentals contribution to mesoscience. This paper presents a detailed discussion on the perspective on this concept.

When a system is open and highly dissipative, as usually found with many situations and regimes in science and engineering, a system's complexity and heterogeneity always present a challenge to describe, which prevail likely in the form of dynamic spatiotemporal structures at the mesoscale in between the microscale of the ``unit" and macroscale of the ``system". This challenge is common in nature and engineering and understanding of many complex systems is blocked at the corresponding mesoscales. Therefore, coarse-graining approaches are typically used to describe such complex systems, even though they tend to lead to unacceptable deviations.

Sometimes, the situations are even worse such that the currently available theories were applied to these complex structures without clarifying their applicability; that is, the variational principles for the systems governed by a single mechanism were mis-applied to the systems governed by at least two dominant mechanisms, leading to confusions and debates. In recent years, the concept of mesoscience was proposed to solve this problem, which is based on the generality of the principle of compromise in competition between dominant mechanisms \cite{book2013,book2014}, called the EMMS principle, which is formulated as multi-objective variational problems \cite{CES2004-1687}. The EMMS principle is believed to reveal the fundamental origin of complexity and diversity across all mesoscales between corresponding unit scales and system scales, and is different from traditional variational approaches. The concept of mesoscience was proposed to verify and extend the possible universality of the EMMS principle \cite{arXiv2009,P2010-634,arXiv2013,book2013,book2014,NSR2017,AR2017}. This paper gives more details on the rationality of mesoscience by case studies.

Many evidence examples have indicated that in many complex systems, mesoscale complexity can be interpreted based on the inherent compromise between competing dominant mechanisms. In the case when only one mechanism dominates, complexity usually does not emerge, so the corresponding state is simple and can be defined using just conservation relationships. However, the situation becomes much more complex - and we would argue, more interesting - when two dominant mechanisms are involved. In these circumstances three regimes can be identified in terms of the relative dominance of these two mechanisms, i.e., $A$-dominated, $A$-$B$ compromising, and $B$-dominated, as illustrated in Fig. 1 \cite{COCE2016-10,E2016-276}.

\begin{widetext}
\begin{figure}[!htb]
\centering
\includegraphics[width=1.0\textwidth]{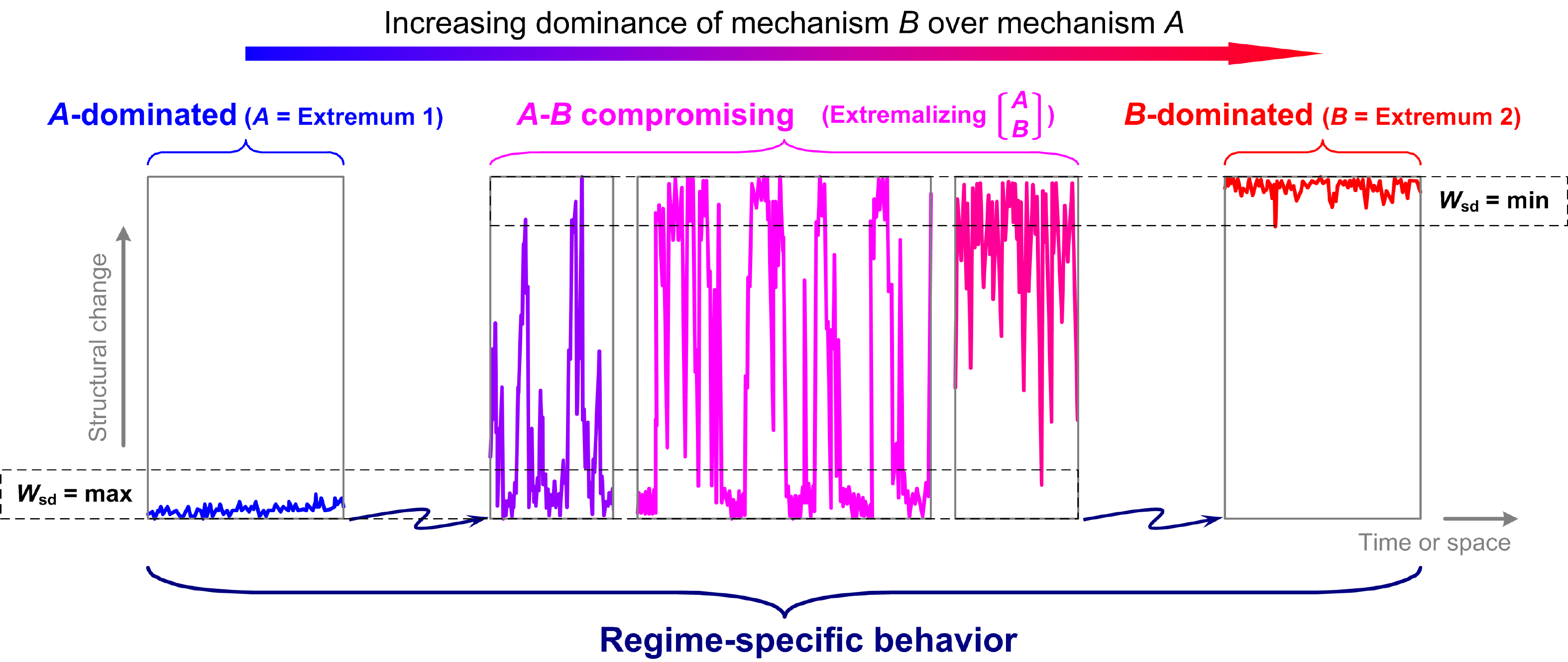}
\label{fig1}
\end{figure}
\vskip\belowcaptionskip
\small\rm
\noindent
FIG. 1: Three regimes that occur successively as the relative dominance of mechanism $B$ over mechanism $A$ changes, and the evolution of structures in the $A$-$B$ compromising regime \cite{COCE2016-10,E2016-276}.
\vskip\abovecaptionskip
\vskip\abovecaptionskip
\vskip\abovecaptionskip
\vskip\abovecaptionskip
\end{widetext}


In the regime with a single dominant mechanism, either $A$ or $B$, a single extremal function can be found, either minimization or maximization. However, description of the $A$-$B$ compromising regime requires a multi-objective variational formulation that integrates the extremal tendencies of its two adjacent regimes, reflecting the compromise between the two competing mechanisms. While this approach is promising to solve many problems related to complexity and diversity in both science and engineering \cite{CEJ2015-112}, its universality needs to be further confirmed; we believe that this should be a key task of mesoscience. On the other hand, the traditional nonequilibrium thermodynamics, both linear and nonlinear, were sometimes used in analyzing mesoscale complex phenomena. However, due to the ignorance of distinguishing regime-specific features of the mesoscale complexity \cite{COCE2016-10,NSR2017,AR2017}, many confusions and debates exist. Meanwhile, there is not much progress made in solving mesoscale problems though great and global attention has been paid to these problems \cite{book2013,book2014}.

On the basis of recent publications \cite{COCE2016-10,CES2016-233,CEJ2016-83}, this paper further gives detailed rationality of mesoscience, gives the reason why the currently available theories are not applicable to formulating mesoscale phenomena, and clarifies the existing confusions and debates in this aspect.

\vfill

\section{Potential Universality: Corroboration by Case Studies}
\label{mysec2}
The generality of the EMMS principle has been preliminarily verified by studying different complex systems with this principle, among which the following are three cases in chemical engineering:

\noindent
\hangafter 1
\hangindent 1.5em
$\bullet$  \textbf{Gas-solid fluidized systems \cite{PT2000-50,COCE2016-10}:} In gas-solid fluidized systems, three typical regimes take place successively, subject to gas velocity $U_\text{g}$ with a specified solid flow rate $U_\text{p}$ (or subject to $U_\text{p}$ with a specified $U_\text{g}$). At low gas velocity, the gas is not capable of suspending solids, therefore, the structure is solid-dominated, which is defined by $\varepsilon = \text{min}$ and/or $W_\text{sd} = \text{max}$, where $\varepsilon$ is the average voidage and $W_\text{sd}$ is the total energy dissipation rate with respect to unit volume of the system; At very high gas velocity, gas possesses sufficient capability to dominate solids in realizing its exclusive dominance defined by $W_\text{st} = \text{min}$, $\varepsilon = \text{max}$, or $W_\text{sd} = \text{min}$, where $W_\text{st}$ is the rate of energy consumption for transporting and suspending particles with respect to unit volume of the system. In this regime, solids are uniformly distributed in the gas flow (in the idealized case) so that the dissipation rate is minimized; However, in between these two regimes, there exists a regime at which neither $W_\text{st} = \text{min}$ nor $\varepsilon = \text{min}$ can dominate or define the structure exclusively. Instead, these two factors have to compromise in their competition between each other, leading to complicated interaction between the state defined by $W_\text{st} = \text{min}$ and that by $\varepsilon = \text{min}$, and resulting in coexistence and alternating appearance of these two different states with respect to time and space. Such compromise can be expressed as a multi-objective variational problem as follows:

\begin{equation}\label{eq1}
    \text {min}
       \begin{bmatrix}
         \varepsilon\\
         W_\text{ st}
       \end{bmatrix}.
\end{equation}

\noindent
\hangafter 0
\hangindent 1.5em
As shown in Fig. 2, also in Fig. 1, $\varepsilon = \text{min}$ defines a very dense structure with the maximized dissipation rate ($W_\text{sd} = \text{max}$), while $W_\text{st} = \text{min}$ gives a very dilute uniform structure with the minimum dissipation rate ($W_\text{sd} = \text{min}$). In the case of compromise in the competition between $W_\text{st} = \text{min}$ and $\varepsilon = \text{min}$, that is, in the compromising regime represented here by $N_\text{st} = W_\text{st}/((1 - \varepsilon)$ $\rho_\text{p}) = \text{min}$ (alternatives are possible according to mathematics) \cite{CES2004-1687}, complicated structures are generated because of the alternating dominance of $W_\text{st} = \text{min}$ and $\varepsilon = \text{min}$ with respect to space and time. In this regime, it should be noted that the interphase interaction between these two states is intensive and was approximately described here through defining a cluster diameter and its volume fraction \cite{book1994}. In this regime, neither minimum nor maximum of total dissipation rate can be used as its variational function. In engineering practice, which regime prevails in a system, or when regime transitions take place, depends on operating conditions, as defined in \cite{FVII1992-83}. The bifurcation between the regimes of $\varepsilon = \text{min}$, $W_\text{st} = \text{min}$ and the compromising regime $N_\text{st} = \text{min}$ is noted at the generalized minimum fluidization velocity \cite{book1994} $(U_\text{mf})_\text{general} = U_\text{mf} + \varepsilon_\text{mf}U_\text{p}/(1 - \varepsilon_\text{mf})$, where (in the case of Fig. 2) the minimum fluidization velocity $U_\text{mf} = 2.34 \times 10^{-3}$ m/s is determined solely by physical properties independent of $U_\text{p}$. The transition from the compromising regime of $N_\text{st} = \text{min}$ to the regime of $W_\text{st} = \text{min}$ is defined by the choking velocity $U_\text{pt}$ at saturation carrying capacity $K^*$. The values of both $(U_\text{mf})_\text{general}$ and $U_\text{pt}$ also depend on the operating conditions.

\begin{widetext}
\begin{center}
\begin{figure}[!htb]
\centering
\includegraphics[width=0.7\textwidth]{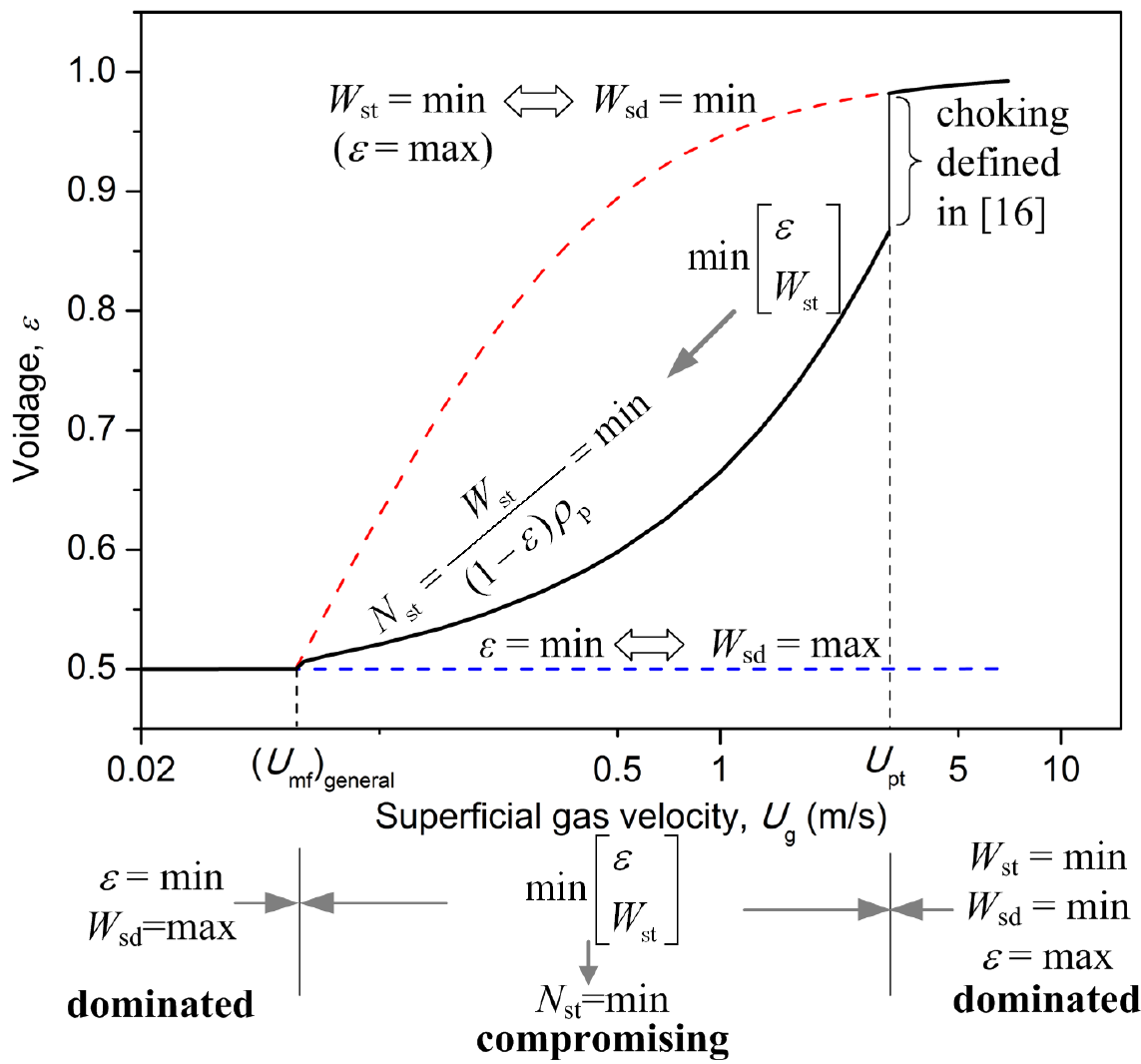}
\label{fig2}
\end{figure}
\end{center}
\vskip\belowcaptionskip
\small\rm
\noindent
FIG. 2: Regime-specific variational features and regime transitions for a gas-solid system ($\rho_\text p$ = 929.5 kg/$\text m^3$, $\rho_\text g$ = 1.1795 kg/$\text m^3$, $d_\text p$ = 54 $\mu\text m$, $\mu_\text g$ = 1.8875$\times 10^{-5} \text { Pa}\cdot\text s$, $\varepsilon_\text {mf}$ = 0.5, $U_\text p$ = 0.0538 kg/($\text m^2\cdot\text s$)).
\vskip\abovecaptionskip
\vskip\abovecaptionskip
\vskip\belowcaptionskip
\vskip\abovecaptionskip
\vskip\abovecaptionskip
\vskip\abovecaptionskip
\vskip\abovecaptionskip
\vskip\abovecaptionskip
\vskip\abovecaptionskip
\vskip\abovecaptionskip
\vskip\abovecaptionskip
\vskip\abovecaptionskip
\vskip\abovecaptionskip
\vskip\belowcaptionskip
\vskip\belowcaptionskip
\vskip\belowcaptionskip
\end{widetext}

\noindent
\hangafter 0
\hangindent 1.5em
In a word, traditional principle of minimum or maximum dissipation rate ($W_\text{sd}$) may hold only for regimes where a single mechanism dominated. However, looking for a single variational function defined by the extremum of total energy dissipation rate is impossible in the $A$-$B$ compromising regime, since the states defined by its minimum ($W_\text{sd} = \text{min}$) and maximum ($W_\text{sd} = \text{max}$) have to compromise in competition. This does not exclude the total energy dissipation rate to be an extremum, but not an absolute extremum without constraint from other dominant mechanisms.

\noindent
\hangafter 0
\hangindent 1.5em
It should be noted that all extrema in the $A$-$B$ compromising regime discussed here are with respect to a volume containing at least one element of mesoscale structure, not for a real local point. This is because it is impossible for either of the two variational functions to be fully realized at the same time at the same local point \cite{CES2004-1687}.


\noindent
\hangafter 1
\hangindent 1.5em
$\bullet$  \textbf{Turbulence:} In a turbulent pipe flow, viscous and inertial effects coexist \cite{CES1999-1151}. The former tends to maintain laminar flow with the minimization of viscous dissipation rate, whilst the latter tends to produce turbulent eddies and maximize the total dissipation rate. In a practical turbulent flow regime, neither the viscous effect nor the inertial effect is realized exclusively; what is actually observed is the result of the compromise between these two competing tendencies. In the regime where the viscous effect dominates, the variational function is $W_\text{v} = \text{min}$ ($W_\text{v}$ is the viscous dissipation rate in unit volume), and in the fully inertia-dominated (probably idealized) regime, it is $W_\text{T} = \text{max}$ or $W_\text{te} = \text{max}$ ($W_\text{T}$ is the total dissipation rate in unit volume, Wte is that for the turbulent (inertial) effect) \cite{CES1999-1151}. In real turbulent pipe flow, that is, the $A$-$B$ compromising regime in Fig. 1, the multi-objective variational problem reflecting the compromise of the above two tendencies can be expressed as (note that $W_\text{T} = \text{max}$ is equivalent to $-W_\text{T} = \text{min}$):

\begin{equation}\label{eq2}
    \text {min}
       \begin{bmatrix}
         W_\text{v}\\
         -W_\text{T}
       \end{bmatrix}.
\end{equation}

\begin{widetext}
\begin{figure}[!htb]
\centering
\includegraphics[width=1.0\textwidth]{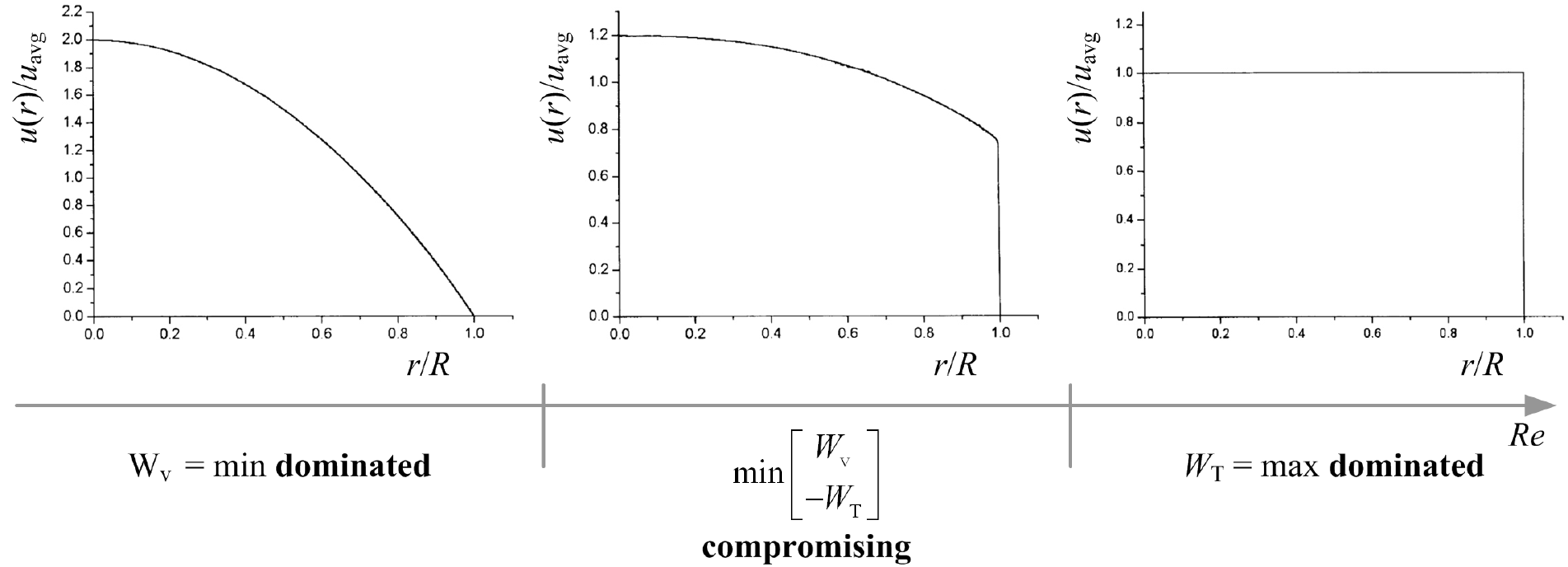}
\label{fig3}
\end{figure}
\vskip\belowcaptionskip
\small\rm
\noindent
\begin{center}
FIG. 3: Illustration of the regime-specific variational features in pipe flow \cite{CES1999-1151}.
\end{center}
\vskip\abovecaptionskip
\vskip\belowcaptionskip
\vskip\belowcaptionskip
\end{widetext}


\noindent
\hangafter 0
\hangindent 1.5em
Once again, a single extremum of dissipation rate cannot truly reflect the variation in the compromising regime. Fig. 3 shows three regimes and their corresponding features, where $R$ is the pipe radius, $u(r)$ the radial velocity, and $Re$ the Reynolds number. Neglecting this kind of compromise between such two dissipative processes leads to the difficulty in revealing the mechanism of turbulence \cite{CEJ2016-83}. This is another evidence showing the compromise in competition between the state with the minimum dissipation rate and that with the maximum dissipation rate.

\noindent
\hangafter 1
\hangindent 1.5em
$\bullet$  \textbf{Heterogeneous catalysis:} For heterogeneous catalysis, we explored the adlayer structures and corresponding processes \cite{CES2016-83}, and identified two competing extremal tendencies reflecting two dominant mechanisms. One is $N_\text{pair}/N_\text{occ} = \text{min}$ ($N_\text{pair}$ is the number of reacting pairs, and $N_\text{occ}$ is the number of occupied sites), reflecting the clustering tendency mainly driven by reactions, which roughly corresponds to the minimization of the total energy dissipation rate. The other is $(N_\text{occ} - N_\text{pair})/N_\text{tot} = \text{min}$ ($N_\text{tot}$ is the total number of sites), which reflects the dispersing tendency basically related to diffusion, adsorption, and desorption processes and corresponds roughly to the maximization of the total energy dissipation rate. In practical cases, usually within the $A$-$B$ compromising regime, neither of these two extreme tendencies can be realized completely. Thus, the heterogeneous structures frequently observed on catalyst surfaces can be viewed as the resulting natural compromise between these two different mechanisms. This multi-objective variational problem can be roughly formulated using these two extremal terms as follows:

\begin{equation}\label{eq3}
    \text {min}
       \begin{bmatrix}[1.5]
         \frac {N_\text { pair}}{N_ \text { occ}}\\
         \frac {N_\text { occ}-N_\text { pair}}{N_ \text { tot}}
       \end{bmatrix}.
\end{equation}

\noindent
\hangafter 0
\hangindent 1.5em
Again, neither maximization nor minimization of the total dissipation rate is dominant exclusively in the compromising regime, showing the alternating appearance between the state of $N_\text{pair}/N_\text{occ} = \text{min}$ and that of ($(N_\text{occ} - N_\text{pair})/N_\text{tot} = \text{min}$, with respect to space and time. This regime-specific feature is illustrated in Fig. 4. Such a feature is of course critical in optimizing catalytic processes, and probably is related to the definition of selectivity \cite{NSR2017}.

\begin{widetext}
\vskip\abovecaptionskip
\begin{figure}[!htb]
\centering
\includegraphics[width=1.0\textwidth]{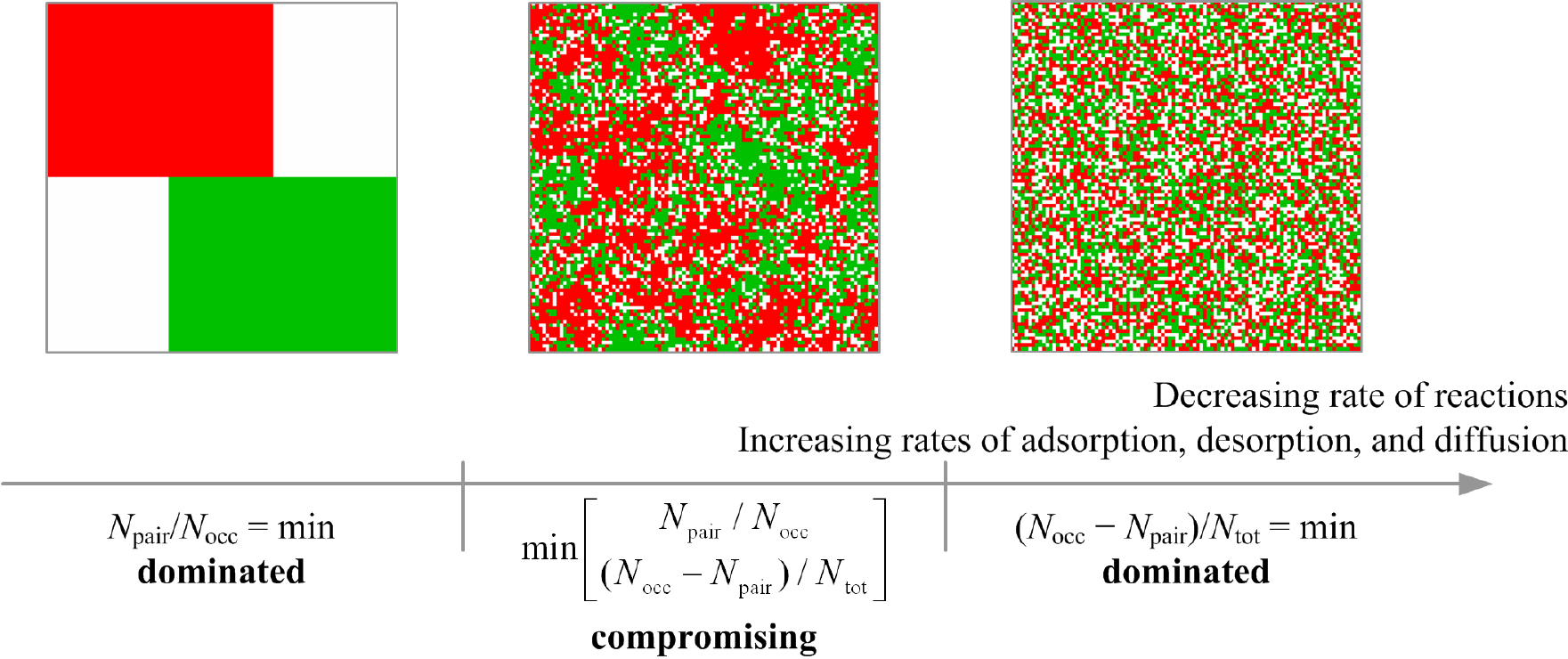}
\label{fig4}
\end{figure}
\vskip\belowcaptionskip
\small\rm
\noindent
FIG. 4: Illustration of the regime-specific variational features in heterogeneous catalysis (different species are distinguished in colors).
\vskip\abovecaptionskip
\vskip\belowcaptionskip
\vskip\belowcaptionskip
\vskip\belowcaptionskip
\vskip\belowcaptionskip
\end{widetext}


In all three cases, it is shown that the EMMS principle is different from the variational approaches that define the variational function of the described systems directly by a single extremum of total energy dissipation rate. The commonality of these three cases can be summarized as follows:

\begin{widetext}
\begin{center}
\begin{figure}[!htb]
\centering
\includegraphics[width=0.8\textwidth]{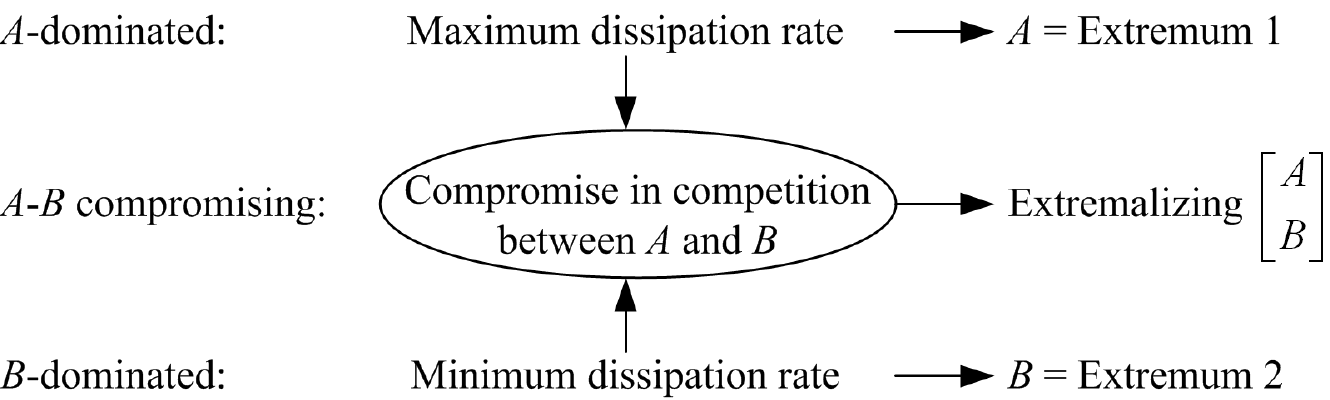}
\end{figure}
\end{center}
\end{widetext}


More evidence on the compromise in competition between minimum and maximum dissipation rates is available, such as metal-nonmetal transition systems \cite{NSR2017}, but needs further detailed analysis.

In this sense, most current variational approaches are valid only for the regimes where a single mechanism dominates; that is, the $A$- or $B$-dominated regime in Fig. 1. However, the EMMS principle bridges these two regimes by considering the compromise in competition between them, which has not yet been tackled by currently available variational approaches. Hopefully, the development of mesoscience will make it possible to remove the existing debate between different approaches, hence, to unify them by complementing this missing link of compromise in competition.

\section{Why Multi-Objective Variational?}
\label{mysec3}
From the EMMS principle, it is clear that two (perhaps more) dominant mechanisms lead to dynamic structures, resulting in rich complexity and diversity \cite{COCE2016-10}. Even for the so-considered equilibrium systems, the EMMS principle also sheds light on, for example, the solid-liquid-gas transition, in which the internal energy compromises with entropy \cite{En2008-462}. In the viewpoint of the EMMS principle, the particular state of solids should be dominated by the minimization of internal energy and that of the gaseous state by the maximization of entropy while the state of liquid should be simply a mixture of ``solids" and ``gas", governed by the compromise in competition between these two extrema, therefore, featuring dynamic changes and flowability. Here, ``solid" means solid-like and ``gas" for gas-like state that might be deviated from the real solid state and that of real gas, respectively, due to the dynamic influence generated in the process of compromise in competition, as shown in Fig. 5. Although the free energy consists of two terms, their compromise in competition, especially, the resulting alternating change between two different states (gas-like and solid-like) has not received sufficient attention. This is due to the inhabited thinking to look for a single extremum of variational functions. That is, traditional understanding of the liquid state is not complete!

\begin{widetext}
\begin{figure}[!htb]
\centering
\includegraphics[width=1.0\textwidth]{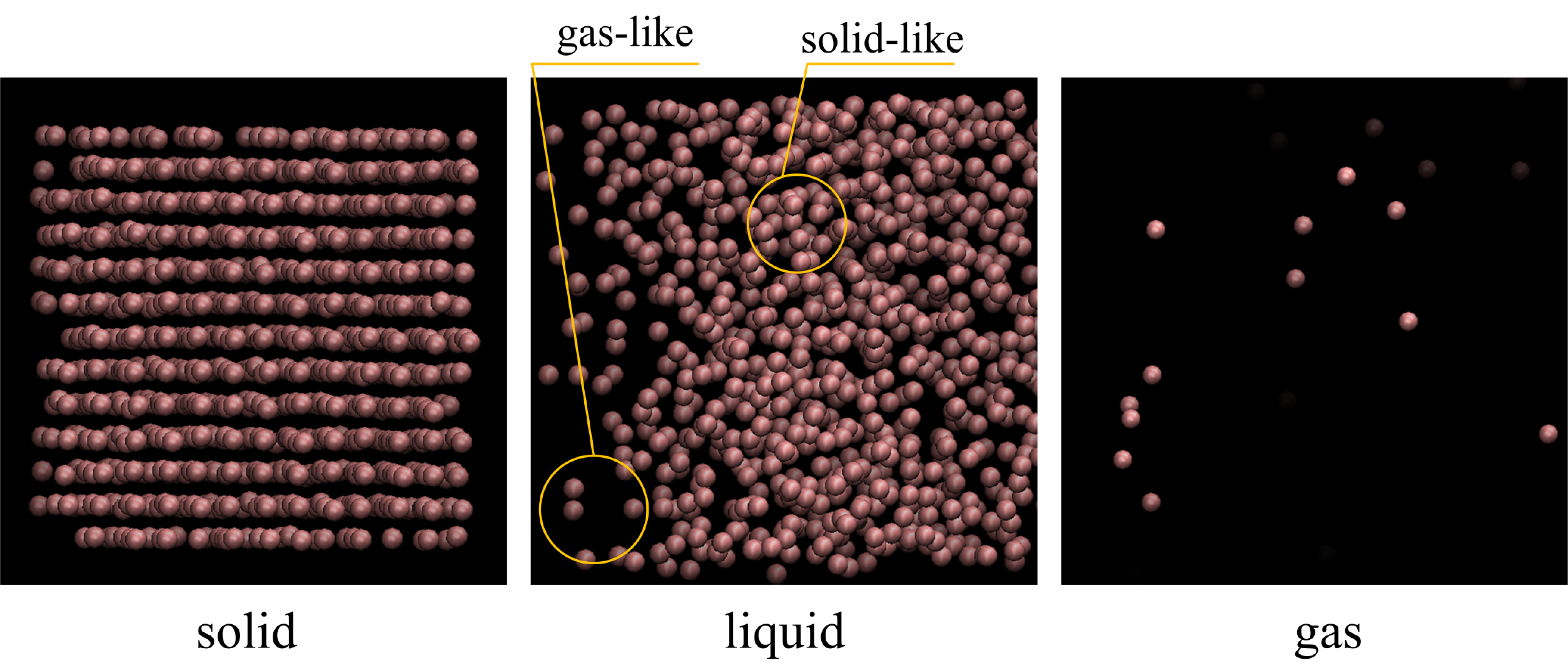}
\label{fig5}
\end{figure}
\small\rm
\noindent
FIG. 5: Liquid as the mixture of solid-like and gas-like states. The results are from molecular dynamics simulation of Argon at 1 atm and 35.9 K (solid), 107.8 K(liquid), and 131.8 K (gas), respectivley, using the 12-6 Lennard-Jones potential with reduced units.
\vskip\abovecaptionskip
\end{widetext}


In traditional studies of complex systems, various variational extremal principles have been proposed, but all consider only a single variational function, leading to much debate around different approaches. We believe that this debate is caused by the fact of the \emph{\textbf{regime-specific}} nature of variational behavior of different dissipative processes. In such an approach, the \emph{\textbf{compromise in competition}} between dominant mechanisms, and the resultant dynamic \emph{\textbf{heterogeneity}} is neglected without distinguishing one mechanism from another \cite{COCE2016-10}, which would show totally different extremal tendencies, that is, regime-specific feature.

\textbf{Regime-specific:} To understand variational features of a system, operating conditions, boundary conditions, and material properties must first be specified, including all exchanges of mass, energy and all interactions occurring at boundaries of the system, all together, called the specified working conditions in this paper (Fig. 6). Then, under these conditions, working regimes should be clarified to know whether the system is in the $A$-dominated, or $A$-$B$ compromising, or B-dominated regime so that we could know the regime-specific variational features. The traditional analysis on ``linear" and ``nonlinear" is not sufficient to distinguish this regime-specific feature. This is a critical issue to be noted.

\setcounter{figure}{5}
\begin{figure}[!htb]
\includegraphics[width=0.45\textwidth]{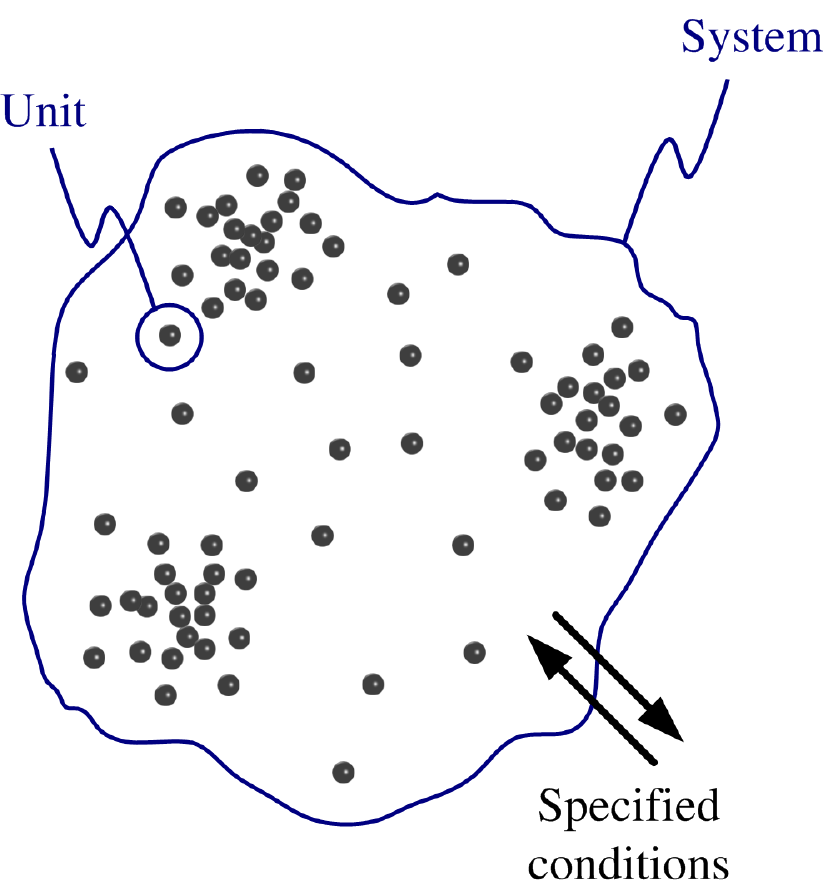}
\caption{\label{fig6} Schematic illustration of a system with the specified conditions. Modified from \cite{book2014}.}
\end{figure}

\textbf{Compromise in Competition:} In fact, a dissipative process could be dominated by either maximization or minimization of dissipation rate or jointly by their compromise in competition, subject to specified working conditions. That is, depending on the specified working conditions, a system can be dominated by totally different dissipative mechanisms or processes showing different variational behaviors, either $A$ or $B$, or more complicatedly and likelier by compromise in competition between $A$ and $B$, as shown in Fig. 1. In particular, the variational properties within the $A$-$B$ compromising regime can be defined neither by maximization nor by minimization of the total energy dissipation rate, as indicated in the above three case studies. This is the reason why attempts to find a single variational function in terms of energy dissipation rate have failed in some cases and minimum and maximum of energy dissipation rate have been confused, leading to debates.

When the specified working conditions are changed, the dominant mechanisms, hence, the dissipative mechanisms in the system, are also changed, leading to regime transitions. When we change the conditions artificially, we must pay attention to the interaction between different dominant mechanisms to distinguish regime transitions caused by the underlying compromise in competition between the two extreme regimes ($A$ and $B$).

Neglecting the compromise in competition between different dominant mechanisms, we argue, leads to confusions between different approaches and the arising of average-based treatments, i.e., neglecting the fundamental heterogeneity from different dissipative processes at the mesoscales. This is a challenging issue in studying complex systems.

\textbf{Heterogeneity:} With the specified working conditions, if there is no heterogeneity within the domain of the system, it is reasonable to assume that the specified conditions hold for every location. This is possible only when one mechanism dominates, i.e., within the $A$- or $B$-dominated regime, where the dissipation rate shows a definite extremal tendency, driven by an identical dissipative mechanism everywhere in the system.

However, if the specified working conditions lead to heterogeneity within the system, the specified conditions should not be assumed to hold everywhere in the system. Actually, in this case, the specified conditions, and hence, the interaction or dissipative mechanisms, can be quite different not only in space but also in time. This is the case particularly and inherently in the $A$-$B$ compromising regime, where the system behavior can not be described with any single dominant mechanism, and the total dissipation rate does not exhibit a single definite and global extremal tendency.

These can be simply detailed a little further. The total energy dissipation rate $\Omega$ for the whole system should be the integration of local density of energy dissipation rate ¦Ø over the whole system, as expressed in Eq.~(\ref{eq4}). When the system is homogeneous, it can be simplified into Eq.~(\ref{eq5}) due to the uniform distribution of dissipation rate density.

\begin{equation}\label{eq4}
 \Omega=\int\limits_{V} \omega\,\text dV
\end{equation}
\begin{equation}\label{eq5}
\Omega=V\omega
\end{equation}

When the system is heterogeneous and the conditions vary in space and in time, different dissipative mechanisms must play roles in the system, alternately in space and time. In this case, $\omega$ is a function of spatial positions $\textbf{r}$ and time $t$, i.e., $\omega(\textbf{r}, t)$, and Eq.~(\ref{eq6}) should not hold any longer, that is, the extremal tendency of $\Omega$ is not equivalent to that of $\omega(\textbf{r}, t)$, as expressed in Eq.~(\ref{eq6}) where Ex($x$) is the extremal tendency of $x$. The behavior of the whole system becomes complicated, and cannot be defined by using a single tendency of energy dissipation rate.

\begin{equation}\label{eq6}
 \text{Ex}(\Omega)|_\text { whole system} \ne \text {Ex}
  \begin{bmatrix}
  \omega(\mathbf {r},t)
  \end{bmatrix}
  |_\text { location \textbf {r}, instant \emph {t}}.
\end{equation}

To explain Eq.~(\ref{eq6}), we present some snapshots taken from a two-dimensional autocatalytic system simulated at different scales in Fig. 7, using both a macroscopic continuum model and the microscopic kinetic Monte Carlo method, respectively. Due to the existence of structural heterogeneity, the specified conditions are distributed in the system, leading to complex changes of different mechanisms that usually correspond to different dissipative processes. Therefore, the total energy dissipation rate of the whole system does not show a definite extremal tendency, as described in our previous work \cite{CES2016-233}. In fact, the energy dissipation rate even within a grid also varies when such heterogeneity is involved, as given in Fig. 7 under the corresponding snapshot of each grid, which was integrated over all the points within the grid. Two dominant mechanisms can be roughly identified, i.e., mechanism $A$ mainly corresponding to diffusion, adsorption, and desorption processes, which tends to homogenize the distribution of species, and mechanism $B$ basically corresponding to reactions, which leads to clustering.

\begin{widetext}
\begin{figure}[!htb]
\centering
\includegraphics[width=1.0\textwidth]{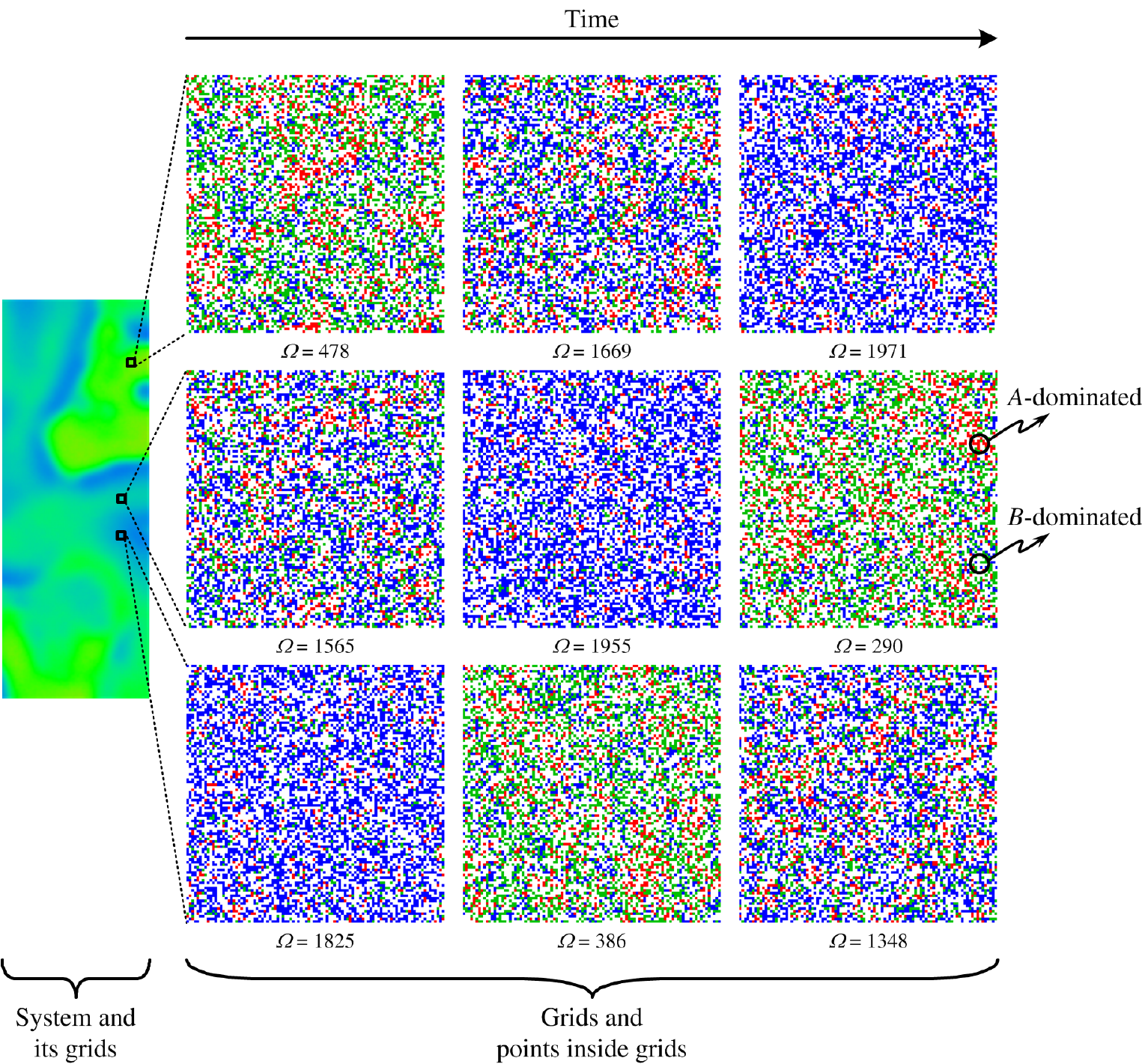}
\label{fig7}
\end{figure}
\vskip\belowcaptionskip
\small\rm
\noindent
\begin{center}
FIG. 7: Illustration of the heterogeneity caused by compromise in competition in a system and its grids.
\end{center}
\vskip\abovecaptionskip
\vskip\belowcaptionskip
\vskip\belowcaptionskip
\vskip\belowcaptionskip
\vskip\belowcaptionskip
\end{widetext}


In other words, as long as dynamic heterogeneity exists, at least two dominant -and competing - mechanisms \emph{must} be involved. In this case, according to the EMMS principle, these mechanisms prevail alternately with respect to space and time. Directly unifying them with a single blurring extremum is impossible, even though an indirect integration of them into a single term is possible, such as $N_\text{st} = \text{min}$ in Fig. 2, but absolutely not in terms of $\Omega$ \cite{COCE2016-10,E2016-276}. This is the reason why there are debates between researchers searching for a single variational function of the total energy dissipation rate.

\textbf{Self-adaptive:} Dynamic processes can be viewed as the responses to the specified working conditions for a system. These responses can play roles in return as a feedback to mediate the specified conditions, as expressed in Fig. 8, the effect of which can be minor at the system scale, but increasingly critical at smaller scales, such as in a simulation grid. A typical example is negative feedback \cite{book2003,JSSMS2016-291}; i.e., the induced processes serve to relax the effect of the specified conditions. In the $A$-$B$ compromising regime, such feedbacks are involved in the spatiotemporal compromise in competition between dominant mechanisms to stabilize the dissipative structures. Therefore, the energy dissipation behavior of the whole system could be complicated.

\setcounter{figure}{7}
\begin{figure}[!htb]
\includegraphics[width=0.45\textwidth]{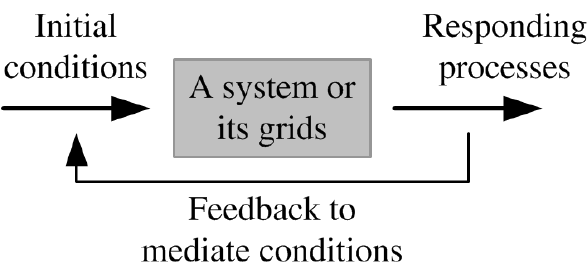}
\caption{\label{fig8} The self-adaptive feature of complex systems makes it impossible to fix the specified conditions everywhere throughout the whole system.}
\end{figure}

In summary, a single variational function may be valid only for systems where identical conditions exist at different locations and instants in time and space. In this case, the extremal tendency of the whole system is consistent with that of any point in it. Such a system is usually and necessarily uniform, operated either in the $A$- or $B$-dominated regime and dominated by only one dissipative mechanism. Even in this case, regime specification, either $A$- or $B$-dominated, needs to be made before defining its variational function since they could be totally different.

Importantly, for most complex systems operated within the $A$-$B$ compromising regime, with the specified conditions, there is not a single variational function in terms of total energy dissipation rate. This arises because there are at least two interaction mechanisms or dissipative processes which can be generated inside the system, and can not be defined either by the one for the $A$-dominated regime or by that in the $B$-dominated regime. Therefore, the multi-objective variational feature has to be considered, and multi-objective variational formulation is necessary. This, we believe, is the core concept of mesoscience. The specified conditions for the system are not now distributed uniformly in the system, but result in multiple dissipative processes which prevail in the system, alternately with respect to time and space, leading to fascinating complexity and diversity. This, we believe, is the missing principle in traditional approaches and complexity science, implying the potential great significance of mesoscience.

\section{Remarks and Perspective}
\label{mysec4}
As summarized in Fig. 9, taking fluidization as an example, real structures in complex systems are spatiotemporally dynamic. In such structures, three different interaction mechanisms prevail, each corresponding to a distinct dissipative process: The structure in the dilute phase is fluid-dominated, while that in the dense phase is particle-dominated. The compromise between these two mechanisms leads to complex and rich interphase interactions. These intrinsic facets give rise to the spatiotemporal dynamic structure that shows alternate appearance of two dominant mechanisms with respect to time and space. Such a governing principle must be considered in any realistic physical model; otherwise, a reasonable solution is simply not possible.

\begin{figure*}[!htb]
\includegraphics[width=1.0\textwidth]{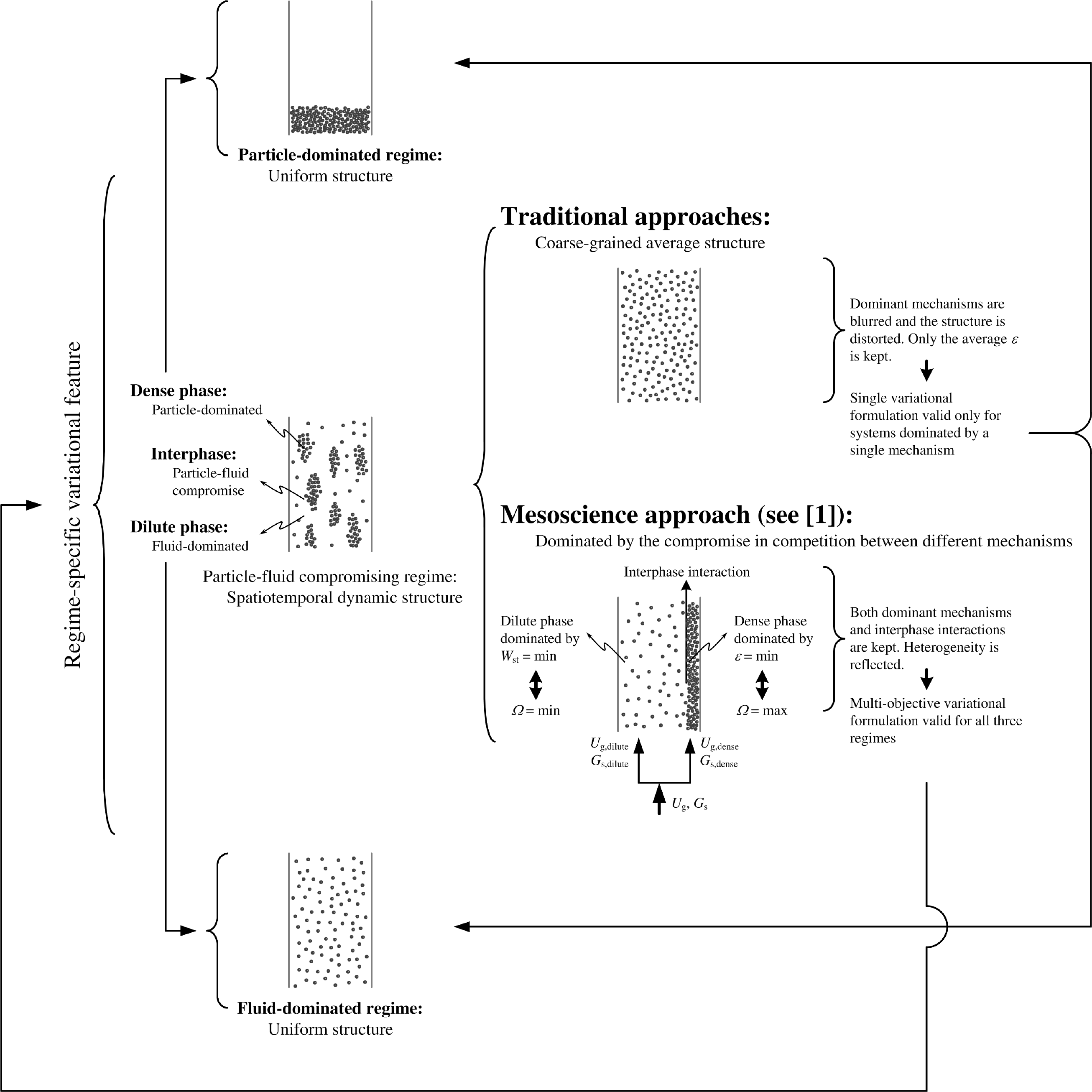}
\caption{\label{fig9} Illustration of regime-specific features and different approaches to describe the compromising regime, exemplified with the gas-solid fluidized system.}
\end{figure*}

Regretfully, these facets have not been taken into account in analyzing complex systems. Because of the complexity and diversity of these structures, such systems are likely analyzed using coarse-graining approaches. The formulations of coarse graining are usually based on experiments and artificial assumptions without considering the compromise in competition between different dominant mechanisms. Therefore, diverse strategies exist in different fields even though they have the common purpose of simplifying the analysis of complex systems. Coarse-graining approaches are usually used either to improve the computational accuracy in average modeling or to lower the computational cost in discrete computation, both of which cause low predictability. Examples of coarse-graining approaches are distribution functions in molecular simulation, constitutive equations in fluid dynamics, and even extremely large-scale averaging in astronomy, to name a few. There is no unified principle for coarse graining, and even worse, some coarse-graining models are not related to multiscale analysis, leading to difficulties when using them to solve engineering problems. That is, the dominant mechanisms are blurred and the structures are distorted when using coarse-graining approaches, as illustrated in Fig. 9. When one tries to identify a variational function for a structure, the approach usually followed is to directly look for a single variational function. However, although this is feasible in the $A$- and $B$-dominated regimes, it becomes impossible, at least difficult, in the $A$-$B$ compromising regime, leading to confusion. This missing principle at mesoscales is currently an important challenge in science and engineering, attention to which is however insufficient.

Mesoscience follows a different approach and is based on the EMMS principle of compromise in competition, as shown in Fig. 9. The heterogeneity in the compromising regime is believed to reflect the compromise in competition between different dominant mechanisms representing different dissipative processes. By distinguishing such distinct mechanisms and their respective structures, the heterogeneity is naturally and accurately described by establishing a multi-objective variational expression for the compromise between competing mechanisms. Since the compromise in competition between dominant mechanisms contains the individual dominance of each mechanism as well, the $A$-$B$ compromising variational formulation can also be simplified to describe the $A$- or $B$-dominated. In other words, it is valid for all three regimes. Actually, the behavior of the $A$- or $B$-dominated regime can be formulated simply using conservation relationships because the system is simple. Of course, the regime transition from $A$-dominated to $A$-$B$ compromising and that from $A$-$B$ compromising to $B$-dominated should be defined. This is another issue that is discussed in ref. \cite{FVII1992-83} and needs to be considered further. Of course, the description of the interphase interaction still represents a fascinating challenge, such as the cluster description in the preceding case study (the first case).

It is evident that traditional coarse-graining approaches lose vital information about compromise-in-competition between two dominant mechanisms, through averaging. For example, a popular treatment of $x = \bar{x} + x^\prime$, i.e., the division of a state variable into its average and fluctuation, which blurs all mechanisms. According to the EMMS principle, a rational simplified coarse-graining method should be expressed as $x = f x_1 + (1 - f) x_2$, where $x_1$ is the state variables defined by mechanism 1 and $x_2$ is that for mechanism 2. Using this logic, many problems not handled by current approaches will find better solutions. The rationale of this prediction was properly verified by the simulation of fluidization, as shown in Fig. 9, where the dilute phase (corresponding to $W_\text{st} = \text{min}$) and the dense phase (corresponding to $\varepsilon = \text{min}$) appear alternately in space and time \cite{CEJ2003-71,CES2007-208}.

We propose that for complex systems, especially for the regime where more than one mechanism dominates, a possible stability criterion should be a multi-objective variational expression that is established on the basis of the EMMS principle. In this regime, no approach looking for a single variational function of total energy dissipation rate is feasible. The EMMS principle represents a bridge between different variational principles. Because of the commonality of heterogeneity and complexity in different systems, this principle or mesoscience deserves to be explored in different fields. In fact, complexity and diversity in the world do not originate from a single mechanism. Coarse-graining approaches must take this fact into account. The EMMS principle is specific to explain complexity and dynamic dissipative processes.

Another missing consideration is the multilevel nature of the natural world. Current approaches have paid insufficient attention to distinguishing levels, so sometimes two or even more levels have been blurred as a single system to study. This causes confusion because different mechanisms exist at different levels \cite{book2014,COCE2016-10}. That is, regime- and level-specific natures are critical to understand the complexity in the world. Of course, this brings about another issue to be addressed - correlation between levels.

In conclusion, the EMMS principle has confirmed the preliminary generality of the compromise in competition between dominant mechanisms. What should be addressed further with the concept of mesoscience have been detailed in \cite{COCE2016-10,NSR2017,AR2017}. In addition to the dissipation rate, the variational functions of dominant mechanisms could be formulated in different terms, which are probably level-specific.

If mesoscience can be established as expected through studying different systems as case studies first, and then examining their commonality \cite{book2013,book2014,CEJ2015-112} from diversity, the current gaps in understanding complex systems will be filled, and many global challenges can be analyzed more rationally. For instance, the success of mesoscience may allow climate change modeling and weather prediction to become even more accurate and faster to enable our capability in sustainable development; or revolutionize rational design and smart manufacture of materials. Mesoscience may also shed light on neuroscience to reveal the secret of cognition, and be integrated with computational science to facilitate artificial intelligence research; and help to understand many as-yet unknown phenomena in electron systems in condensed matter, and even the uncertainty principle in quantum mechanics. Thus, establishment of mesoscience deserves global effort and joint action from the whole spectrum of science and support from our whole society. Fortunately, the National Natural Science Foundation of China (NSFC) has launched a mesoscience program focusing on the two levels of mesoscales in chemical engineering \cite{SSC2014-277}.

Concerning the traditional thermodynamics, particularly the extremalization principles of dissipation, formulated by $\omega = \sum\limits_{i} X_{i}J_{i}$, we would conclude that they are applicable to the two extreme regimes dominated only by a single dissipative mechanism, that is, $\omega = \sum\limits_{i} X_{i}J_{i} = \text{min or max}$ under $A$ or $B$, respectively. However, the extremalization of this expression with a single objective is not applicable to the $A$-$B$ compromising regime with multi-objectives. This is the core concept of mesoscience.

The opinions expressed in this article are somewhat premature, potentially provocative and need to be verified. Even for the EMMS model itself, many details are to be refined, such as the interphase interaction. However, we are encouraged by the exciting clues of generality - potentially, even universality - in this direction which emerges in case studies, which is far beyond our capability to effectively explore and to reach a final conclusion. Therefore, we write this article to call on wide attention and to welcome interest and criticism from many different disciplines.

\begin{acknowledgments}
We would thank critical comments from Profs. Ying Hu, Quan Yuan, Mingyuan He, Lei Guo, and Peter Edwards. We appreciate the financial support from the National Natural Science Foundation of China (Grant NOs. 91334000, 91534104, and 91334102), the Research Center for Mesoscience at the Institute of Process Engineering, Chinese Academy of Sciences (Grant No. COM2015A002), and the State Key Laboratory of Multiphase Complex Systems (Grants No. MPCS-2015-A-03).
\end{acknowledgments}

%

\end{document}